# Digital predistortion for power amplifiers using separable functions

Hong Jiang and Paul Wilford
Bell Laboratories, Alcatel-Lucent
700 Mountain Ave, Murray Hill, NJ 07974-0636
{hong.jiang, paw}@alcatel-lucent.com

*Abstract*— This paper is concerned with digital predistortion for linearization of RF high power amplifiers (HPAs). It has two objectives. First, we establish a theoretical framework for a generic predistorter system, and show that if a postdistorter exists, then it is also a predistorter, and therefore, the predistorter and postdistorter are equivalent. This justifies the indirect learning methods for a large class of HPAs. Secondly, we establish a systematic and general structure for a predistorter that is capable of compensating nonlinearity for a large variety of HPAs. This systematic structure is derived using approximation by separable functions, and avoids selection of predistorters based on the assumption of HPA models traditionally done in the literature.

## I. INTRODUCTION

Linearization of RF high power amplifiers (HPAs) using digital predistortion has been widely studied in the literature and has been routinely implemented in telecommunication and broadcast systems. In a digital predistortion system, the transmitted signal is processed by a predistorter in the digital domain before it is converted to analog, upconverted to RF, and amplified. The purpose of the predistorter is to compensate nonlinear effects of the HPA, which include distorting signal constellation and spreading signal spectrum. Consequently, the use of digital predistortion can increase the efficiency of HPAs by transmitting at high output power without suffering from undesired nonlinear effects that impact the system performance.

There are many techniques to perform digital predistortions, see [1]-[5] and references therein. They are all based on the fundamental idea that a predistorter is constructed to compensate the nonlinear effects of the function representing the HPA. Then the cascade of the predistorter and the amplifier approximates the linear or identity operator, and hence the nonlinearity of the amplifier is removed or reduced. Predistorters have been constructed by using memory polynomials [1] [2], full Volterra representations [4], or a neural network [5]. These techniques are justified by studies showing that HPAs may be reasonably represented by Volterra or memory polynomial models [6]-[8].

In a predistortion algorithm, the choice of predistorter structure and its computation are critical. A direct learning method [3] computes the predistorter directly such that the predistorter function followed by the HPA function is the identity function. An indirect learning method [1] [2] computes a postdistorter, which is a function preceded by the HPA function. Then a copy of the postdistorter is used as the predistorter. There have been concerns [3] as to whether a postdistorter can act as a predistorter because the HPA function may not commute with the postdistorter. This was addressed in [1] and [14] for the Volterra HPA models. In [14], the necessary and sufficient conditions for the existence of $p$th-order inverse of a nonlinear system of Volterra series were obtained, and it was shown that the $p$th-order post-inverse and pre-inverse are equivalent. Thus, the $p$th-order inverse can be used as either a postdistorter or a predistorter interchangeably for an HPA of Volterra series. However, the question still remains, if a postdistorter is not the $p$th-order inverse, can the postdistorter be used as a predistorter, even for HPAs of Volterra series? More discussions will be given in Section II.

In both the direct learning and indirect learning methods, the pre- or post-distorters are assumed to have certain structures. Traditionally, these structures are constructed according to the assumption on HPA models. The structures may need to be modified and optimized with different choices of HPA models.

This paper has two objectives. First, we establish a theoretical framework for a generic predistorter system, and show that for any HPA, whether or not it can be expressed by a Volterra series, if a postdistorter exists, then it is also a predistorter, and therefore the postdistorter and predistorter are equivalent. This justifies the indirect learning methods for a large class of HPAs, because in an indirect learning method, the goal is to find a postdistorter, and once found, it is used as the predistorter. This framework also leads to a method for constructing a postdistorter for any HPA without knowing its characteristics. Secondly, we establish a systematic and general structure for a predistorter that is capable of compensating nonlinearity for a large variety of HPAs. This systematic structure is derived independent of HPA models by using approximation by separable functions, and avoids the selection of predistorters according to the assumption of the HPA model traditionally done in the literature. An advantage of this systematic structure is that the same structure can be used to linearize a large variety of HPAs. Therefore, the same algorithm can work effectively without needing changes in different applications with different types of HPAs.

The paper is organized as follows. In Section II, a theoretical framework is established where we show the equivalence of the predistorter and postdistorter, if a postdistorter exists. In Section III, we present a systematic and general structure for a predistorter. The structure is derived from the approximation of a multivariate function by a sum of separable functions. Simulations are given in Section IV to demonstrate the effectiveness of the separable function predistorters. Finally, conclusions are given in Section V.

## II. THEORETICAL FORMULATION

In this section, we formally define the predistorter and postdistorter for HPAs with memory effects, and show that for any HPA, if a postdistorter exists, it is also a predistorter. We also demonstrate how a postdistorter can be constructed without knowing the structure of HPA, which motivates the discussions of section III.

A generic digital predistortion system is shown in Figure 1. Here $x_n$ are complex samples of the baseband signal in the time domain, $n$ is the time sample index, $z_n$ are complex samples after the predistorter, and $y_n$ are the digitized complex samples of the signal after the HPA. The HPA is assumed to have memory effects. Since we are interested in digital processing, we combine the processing of DAC, upconverter, HPA, feedback, downconverters and ADC into one function $F$. That is, $F$ is an operator mapping the sequence of complex numbers $\{z_n, n = 0,1,...\}$ to the sequence of complex numbers $\{y_n, n = 0,1,...\}$ shown in Figure 1. We will informally call $F$ the transfer function of the HPA, and are interested in predistortion of this function.

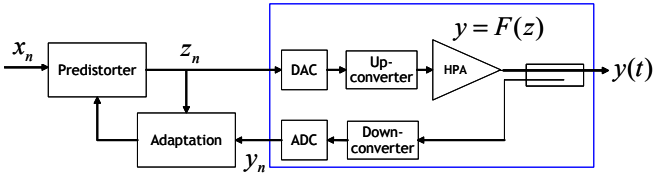

Figure 1. Generic digital predistortion

Let $C^{\aleph_0}$ be the space of all infinite sequences of complex numbers. Formally, we define the function representing an HPA with memory effects to be an operator defined on $C^{\aleph_0}$, i.e., $F : C^{\aleph_0} \to C^{\aleph_0}$. Let $y = \{y_n, n = 1,2,...\} \in C^{\aleph_0}$, $z = \{z_n, n = 1,2,...\} \in C^{\aleph_0}$. Then we denote the mapping $F : z \mapsto y$ by $y = F(z)$. We say that the gain of $F$ is $G$ if $|z_n| = 1$, for all $n$, implies $|y_n| = G$ for all $n$, where $y = F(z)$. We say that $F$ has unit gain if $G = 1$. Without loss of generality, we can assume that the HPA function $F$ has unit gain. This is because, if necessary, we can always adjust the power of the feedback signal, and the bit width of the ADC so that the digital feedback signal $y_n$ has expected value equal to 1. Let $\mathcal{D}(F) \subset C^{\aleph_0}$ and $\mathcal{R}(F) \subset C^{\aleph_0}$ be the domain and the range of operator $F$, respectively.

For a given complex-valued function of $Q$ complex variables $P : C^Q \to C$, we can define an operator $P : C^{\aleph_0} \to C^{\aleph_0}$. Let $x = \{x_n, n = 1,2,...\} \in C^{\aleph_0}$, so that $z = P(x) = \{z_n\} \in C^{\aleph_0}$ is defined by the infinite sequence

$$z_n = P(x_n, x_{n+1}, ..., x_{n+Q-1}), n = 1, 2, .... \quad (1)$$

The operator $P : C^{\aleph_0} \to C^{\aleph_0}$ of (1) is said to be induced from the multivariate function $P : C^Q \to C$. Since there will be no confusion, we will use the same notation for both multivariate function and the operator on $C^{\aleph_0}$ induced from it. Some remarks regarding the notation in (1) will be given at the end of this section.

We can now define the predistorter for $F$. An operator $P_r : \mathcal{D}(F) \to \mathcal{D}(F)$ induced by a complex-valued function of $Q$ complex variables $P_r : C^Q \to C$ is said to be a predistorter for $F$ with memory depth of $Q$, if for every $x \in \mathcal{D}(F)$, the operator $P_r(x)$, as defined by (1), satisfies $P_r(x) \in \mathcal{D}(F)$, and

$$F[P_r(x)] = x, \text{ for all } x \in \mathcal{D}(F). \quad (2)$$

Referring to Figure 1, if $P_r$ is a predistorter for the HPA function $F$, then we have $y = F(z) = F[P_r(x)] = x$. This shows that the cascade of the predistorter and the HPA transfer function results in the identity function, and therefore, the linearization of the HPA is achieved by the predistortion. Again, we will refer both the operator $P_r : \mathcal{D}(F) \to \mathcal{D}(F)$, and the multivariate function $P_r : C^Q \to C$ as predistorter. We use subscript $r$ to signify that the operator is a predistorter ($P_r$).

In advanced digital predistortion algorithms, an adaptive method is used to compute the predistorter. This is represented by the adaptation block in Figure 1. Different predistortion techniques use different methods for computing the predistorter. The performance of a digital predistortion algorithm is largely determined by the structure of the predistorter and how well the predistorter is computed. It is therefore critical to have some practical ways to construct the predistorter.

Although it is not immediately obvious how to construct a predistorter, there is a related function that is implicitly defined by the HPA operator $F$, as illustrated by Figure 2. Consider an input/output pair $z, y$ of $F$, and assume that the pair can be considered as an output/input pair of an operator. Although this operator is not explicitly known, its input and output can be observed through the action of $F$. If this operator can be induced from a multivariate function, then it is called a postdistorter.

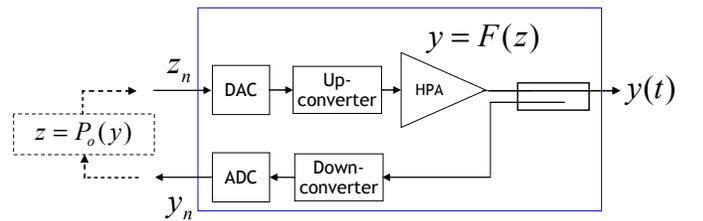

Figure 2. Illustration for the definition of postdistorter

An operator $P_o : \Re(F) \to \mathcal{D}(F)$ induced by a complex valued function of $Q$ complex variables $P_o : C^Q \to C$ is said to be a postdistorter of $F$ with memory depth of $Q$, if the action of $P_o$ on the output of $F$ results in its input, i.e.,

$$z = P_o[F(z)], \text{ for all } z \in \mathcal{D}(F). \quad (3)$$

We use subscript $o$ to signify that the operator is a postdistorter ($P_o$).

It can be easily shown that if a postdistorter exists, then $F$ must be a bijection from $\mathcal{D}(F)$ onto $\Re(F)$. Indeed, let $z_1, z_2 \in \mathcal{D}(F)$, and $F(z_1) = F(z_2)$. If $P_o$ is a postdistorter, then we have $z_1 = P_o[F(z_1)] = P_o[F(z_2)] = z_2$ according to (3), which shows that $F$ is a bijection. Now since $F$ is a bijection, it has a unique inverse. The unique inverse must be equal to the postdistorter and it is also the predistorter. This proves the following property.

**Property 2.1.**

For any HPA, if a postdistorter $P_o$ exists, then there is a unique predistorter $P_r$, and the predistorter is equal to the postdistorter, i.e., $P_r = P_o$.

Property 2.1 justifies the indirect learning methods: if we can find a postdistorter for the HPA, then it is the predistorter. An advantage of the postdistorter is that the definition of the postdistorter itself suggests a way to compute it. To construct the postdistorter, we make use of the HPA operator $F$. We regard the input of the HPA as the output of the postdistorter, and the output of the HPA as the input of the postdistorter. In other words, the input and output of the postdistorter can be observed by observing the output and input of the HPA, respectively. Observations of the input and output of the HPA enable us to construct the postdistorter without knowing the internal structure or characteristics of the HPA.

It is worthwhile to point out that Property 2.1 is not symmetric with respect to the predistorter and postdistorter. That is, if a predistorter exists for an HPA, it does not necessarily follow that a postdistorter also exists, because the existence of a predistorter does not guarantee bijection of HPA.

In comparison to the result of [14], Property 2.1 applies to a broader range of HPAs. The condition of Property 2.1 that a postdistorter exists is hardly a real limitation on HPAs that are linearizable because issues of saturation and dead zone related to HPAs can be usually dealt with using other techniques such as the crest factor reduction. On the other hand, the result of [14] is stronger than Property 2.1 for HPAs that can be expressed as Volterra series; it provides necessary and sufficient conditions for the existence of the $p$th-order inverse for those HPAs. However, the $p$th-order inverse may not necessarily be a predistorter defined in the sense of (2), because a $p$th-order inverse may not be an inverse (see Equation (5) of [14]). Although the $p$th-order inverse can be used as a predistorter for an HPA of Volterra series because it can be a good approximation of the inverse, a good predistorter does not have to be a $p$th-order inverse, even for HPAs of Volterra series. One may be able to construct a better approximation of the inverse than the $p$th-order inverse by some other means. The result of [14] does not address whether a postdistorter that is not the $p$th-order inverse can be used as the predistorter, which is exactly the gap that Property 2.1 now fills. Indeed, in this paper, we will construct postdistorters that may not be $p$th-order inverses.

Some remarks regarding the notation in (1) are in order. The operator $P$ induced from a multivariate function is defined by (1) instead of the conventional form

$$z_n = P(x_n, x_{n-1}, ..., x_{n-Q+1}), n = Q, Q+1, .... \quad (4)$$

This is done purely for convenience to avoid dealing with initial conditions for $n < Q$, because (1) is defined for all $n$, but (4) is defined only for $n \geq Q$. There are no fundamental differences between them, if the first $Q-1$ terms, $z_1, z_2, ..., z_{Q-1}$, are removed from (4) and the index for $z$ is renamed from $n$ to $n-Q+1$. Consequently, we may use either (1) or (4) for the operator induced from a given multivariate function. In actual implementations, it is more appropriate to use the conventional form of (4). In particular, in the following sections, we will switch to using the conventional form of (4) instead.

We conclude this section with a note about the approximation of pre- or post-distorter. When no postdistorter of a finite memory depth exists, we will naturally attempt to find an approximation by using a function with a finite memory depth. With appropriate conditions, such as a condition on the continuity placed on the HPA function, it is reasonably expected that a good enough approximation of the postdistorter is also a good approximation of the predistorter. Also, when the input signal $x_n$ is not in the range of the HPA function $F$, $\Re(F)$, because, for example, no output of the HPA can be made to match the input signal, an approximation of the input signal is needed to properly define what is meant by linearization. Theoretical treatment of the approximation has not been considered here, but can be dealt with by placing continuity conditions on the HPA function, the predistorter, and the postdistorter.

### III. PREDISTORTER WITH SEPARABLE FUNCTIONS

In the previous section, we have established that if we can find a postdistorter induced from a multivariate function, then it is also the predistorter for the HPA. Furthermore, in order to find a postdistorter, all we need to do is to construct a multivariate function that matches the input and output of the HPA. In this section, we discuss how to construct multivariate functions that approximate the postdistorter for a generic HPA.

Two major functional blocks are present in a digital predistortion system as shown in Figure 1. The predistorter block is straightforward; it computes the output sequence $z$ from the input sequence $x$ for the given predistorter. The adaptation block requires more intelligence; its responsibility is to construct the predistorter. In this section, we discuss implementation issues for both of these blocks.

As shown in the previous section, a predistorter, if it exists, is the same as the postdistorter, which can be computed with the aid of the HPA function by observing its input and output. Therefore, the construction of the predistorter becomes a multivariate regression problem. More specifically, we take samples from the input and output of the HPA, $z = \{z_1, z_2, ..., z_N\}$, $y = \{y_1, y_2, ..., y_N\}$, with $y = F(z)$. We then look for a multivariate function $P: C^Q \to C$ that best matches the output and input samples taken from the HPA. That is, we look for $P$ so that the finite sequence $\{P(y_n, y_{n-1}, ..., y_{n-Q+1}), n = Q, Q+1, ..., N\}$ best approximates $\{z_Q, z_{Q+1}, ..., z_N\}$. Note that we have used the conventional form of (4) to define the induced operator from a multivariate function, which we will continue for the rest of this paper. Such problems have been studied previously; see [9], [10] and references therein. Due to the curse of dimensionality [9], the complexity of the multivariate regression quickly becomes unmanageable as the number of dimensions, the memory depth $Q$ in our case, gets moderately large, say, $Q \geq 3$.

It has been known [11] that a multivariate function may be approximated by a sum of separable functions. Specifically, the multivariate postdistorter can be approximated by the following

$$P_o(x_n, ..., x_{n-Q+1}) \approx P(x_n, ..., x_{n-Q+1}) = \sum_{k=1}^{K} \prod_{q=1}^{Q} P_{kq}(x_{n-q+1}). \quad (5)$$

Then we can choose the right hand side of (5) as a predistorter, and the induced operator $\tilde{P}: x \mapsto z$ is given by

$$z_n = P(x_n, x_{n-1}, ..., x_{n-Q+1}) = \sum_{k=1}^{K} \prod_{q=1}^{Q} P_{kq}(x_{n-q+1}). \quad (6)$$

Note that the expression on the right hand side of (5) or (6) is not unique; different expressions may represent the same function. In actual implementations, the computed functions $P_{kq}$ depend on the structures chosen and the algorithms used in the computation. It is a good practice to arrange the structures of $P_{kq}$ so that the terms in (6) are distinct, as is done in (10) below.

A predistorter of the form (6) is easy to implement in a hardware platform such as FPGA and ASIC. A schematic of this predistorter is shown in Figure 3.

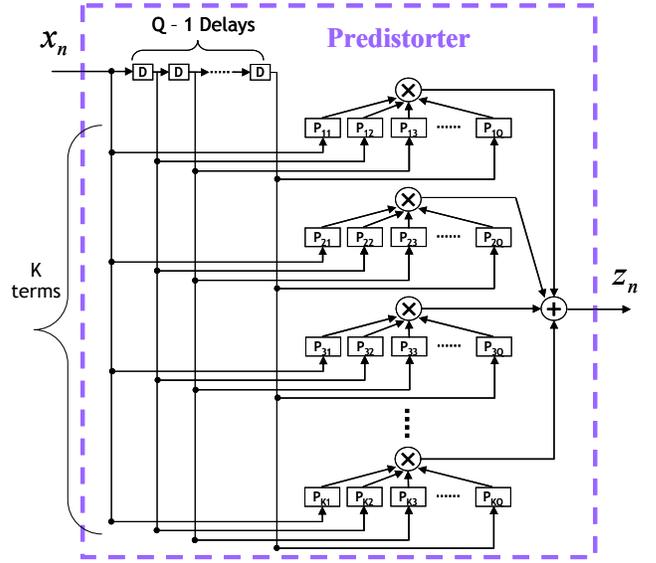

Figure 3. Predistorter using separable functions

The functions $P_{kq}$, $k = 1, ..., K$, $q = 1, ..., Q$, in (6) form a $K \times Q$ matrix of functions, each of which is a complex-valued function of one complex variable.

Now the construction of the predistorter becomes a problem of finding the predistorter matrix, i.e., $KQ$ functions $P_{kq}$, $k = 1, ..., K$, $q = 1, ..., Q$, so that (6) best approximates the observed samples taken from the HPA. More precisely, we take $N$ input and output samples from the HPA, $z = \{z_1, z_2, ..., z_N\}$, $y = \{y_1, y_2, ..., y_N\}$, with $y = F(z)$. We look for functions $P_{kq}$ to minimize the error

$$e_n = z_n - \sum_{k=1}^{K} \prod_{q=1}^{Q} P_{kq}(y_{n-q+1}), n = Q, ..., N. \quad (7)$$

In other words, we look for functions $P_{kq}$ so that

$$\sum_{n=Q}^{N} e_n^* e_n \\ = \sum_{n=Q}^{N} \left[ z_n - \sum_{k=1}^{K} \prod_{q=1}^{Q} P_{kq}(y_{n-q+1}) \right]^* \left[ z_n - \sum_{k=1}^{K} \prod_{q=1}^{Q} P_{kq}(y_{n-q+1}) \right] \quad (8)$$

is minimized. Each function $P_{kq}$ in (8) can be represented by a polynomial, or a linear combination of some known basis functions. Methods for finding the solutions to (8) can be found, e.g., in [10] and [12]. In [10], the solution is obtained by an iterative method in which a linear equation is solved at each iteration to compute the coefficients for the basis function of $P_{kq}$. In [12], each $P_{kq}$ is expressed as a linear combination of orthonormal basis functions, and it found by a stochastic conjugate gradient method. It is also possible to represent each $P_{kq}$ using a look-up table (LUT) [13], and the entries of the LUTs can be computed directly by solving the minimization problem (8).

With the separable functions in (5), the complexity may still be high for implementation on certain platforms, since the functions $P_{kq}$ are functions of one complex variable, and hence two real variables. Further simplifications are possible. It has been demonstrated in [1] and [2] that it is reasonable to represent the nonlinearity in terms of the envelope of the samples. More specifically, for each $k$ in the summation of (6), we pick a factor in the product, say, $m_k$ with $1 \le m_k \le Q$, and rewrite $P_{km_k}(x^{(m_k)})$ as $x^{(m_k)} P_{km_k}(|x^{(m_k)}|)$. For the rest of factors in the product for which $q \ne m_k$, we replace the argument of $P_{kq}(x^{(q)})$ by its amplitude to get $P_{kq}(|x^{(q)}|)$. In this way, we simplify (6) to

$$z_n = \sum_{k=1}^{K} x_{n-m_k+1} \prod_{q=1}^{Q} P_{kq}(|x_{n-q+1}|)$$
$$= x_{n-m_1+1} P_{11}(|x_n|) P_{12}(|x_{n-1}|)...P_{1Q}(|x_{n-Q+1}|)$$
$$+ x_{n-m_2+1} P_{21}(|x_n|) P_{22}(|x_{n-1}|)...P_{2Q}(|x_{n-Q+1}|) \quad (10)$$
$$+ ...$$
$$+ x_{n-m_K+1} P_{K1}(|x_n|) P_{K2}(|x_{n-1}|)...P_{KQ}(|x_{n-Q+1}|).$$

where $1 \le m_k \le Q$.

The advantage of (10) over (6) is that each of the functions $P_{kq}$ is now a function of one real variable $|x_{n-q+1}|$, which makes the implementation simpler. Again, each of the functions $P_{kq}$ may be constructed by a polynomial, a linear combination of some known basis functions, or a LUT.

Equation (10) represents a systematic and general structure for predistorters without assumptions on the HPA model other than the assumption that (6) may be replaced by (10). This structure is amenable to hardware implementation, and is flexible to be adapted with different levels of complexity. The parameters that affect the complexity include the number of the terms in the summation $K$, the memory depth $Q$, the attributes of the functions $P_{kq}$, e.g., degrees of polynomial, and the number of LUT entries. The complexity of the implementation can be further reduced by limiting the number of active functions by setting some of the entries in the predistorter matrix equal to 1, i.e., $P_{kq} \equiv 1$ for some $k,q$.

One reduced form of (10) is derived when the number of summations and the memory depth are equal. In that case, we set $K = Q$ and $m_k = k$. Then, the induced operator from (10) is given by

$$z_n = \sum_{k=1}^{Q} x_{n-k+1} \prod_{q=1}^{Q} P_{kq}(|x_{n-q+1}|)$$
$$= x_n P_{11}(|x_n|) P_{12}(|x_{n-1}|)...P_{1Q}(|x_{n-Q+1}|)$$
$$+ x_{n-1} P_{21}(|x_n|) P_{22}(|x_{n-1}|)...P_{2Q}(|x_{n-Q+1}|) \quad (11)$$
$$+ ...$$
$$+ x_{n-Q+1} P_{Q1}(|x_n|) P_{Q2}(|x_{n-1}|)...P_{QQ}(|x_{n-Q+1}|).$$

A schematic showing an implementation of this predistorter is shown in Figure 4.

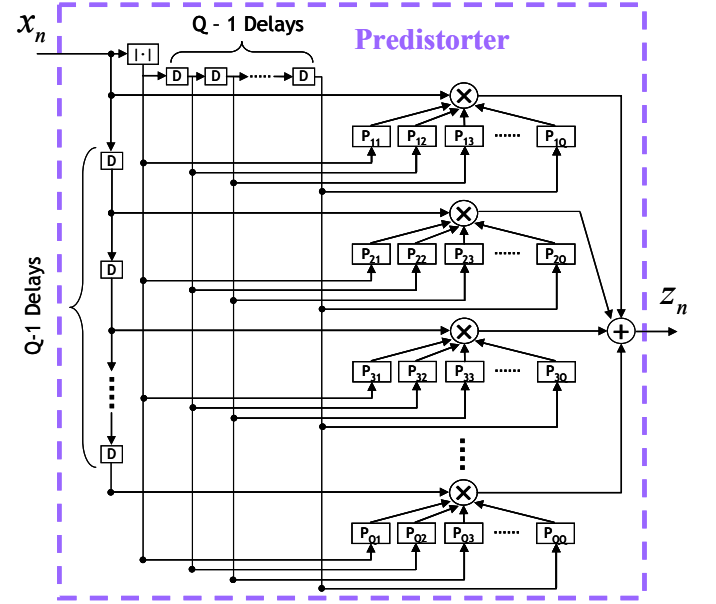

Figure 4. Predistorter of Equation (11)

The memory polynomial predistorter of [2] is a special case of (11), with the off-diagonal entries of the predistorter matrix being the constant 1, and the diagonal entries being polynomials, i.e., $P_{kq}(|x_{n-q+1}|) \equiv 1$ for $k \ne q$, and $P_{kk}$ being a polynomial for each $k = 1, ..., Q$.

Another reduced form of (10) is derived when $K = Q^2$. In this case, we define $m_k = [(k-1) \mod Q] + 1$ and $P_{kq} \equiv 1$ for $q \ne [(k-1) \mod Q] + 1$. Also, $P_{kq}$ is renamed $P_{pq}$ if $k = (p-1)Q + q$. Then the induced operator in this case is given by

$$z_n = \sum_{k=1}^{Q} x_{n-k+1} \sum_{q=1}^{Q} P_{kq}(|x_{n-q+1}|)$$
$$= x_n [P_{11}(|x_n|) + ... + P_{1Q}(|x_{n-Q+1}|)]$$
$$+ x_{n-1} [P_{21}(|x_n|) + ... + P_{2Q}(|x_{n-Q+1}|)] \quad (12)$$
$$+ ...$$
$$+ x_{n-Q+1} [P_{Q1}(|x_n|) + ... + P_{QQ}(|x_{n-Q+1}|)].$$

A schematic showing an implementation for (12) is shown in Figure 5.

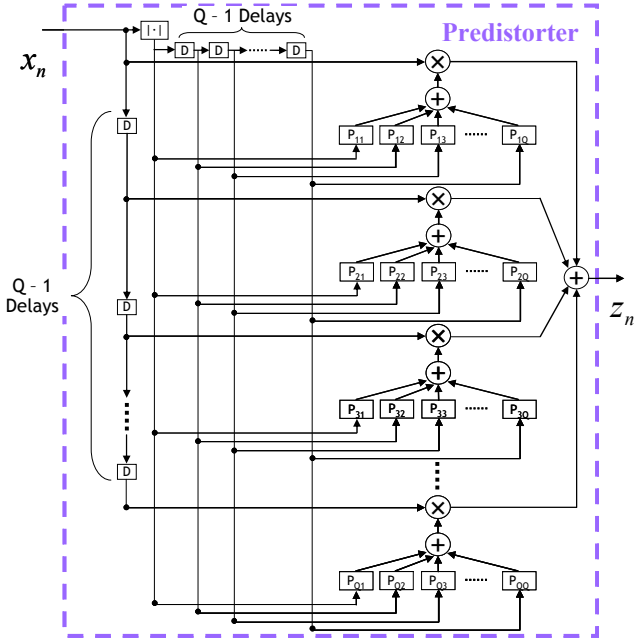

Figure 5. Predistorter of (12)

The generalized memory polynomial predistorter of [1] containing cross-terms is a special case of (12) with each $P_{kq}$ being a polynomial.

Equation (11) is more preferred than (12) because the former has more multiplicative cross-terms while there are only additive cross-terms in the later. For example, (12) lacks the cross terms such as $z_{n-2}|z_{n-1}||z_n|$. Simulation results in the next section will show that there are HPA models that can be linearized by (11), but cannot be linearized by (12). An advantage of (12) is that when the functions $P_{kq}$ are chosen as polynomials or linear combinations of known basis functions, the coefficients of $P_{kq}$ can be computed by solving a system of linear equations. However, even if (11) is used, the coefficients can still be determined by an iterative method in which a linear system is solved at each iteration, see, e.g., [9] and [10].

For a given memory depth $Q$, (11) and (12) can be further simplified by setting some entries in the predistorter matrix to identify (in Equation (11)) or zero (in Equation (12)), i.e., $P_{kq} \equiv 1$, for some $k, q$ in (11) and $P_{kq} \equiv 0$, for some $k, q$ in (12).

We end this section with a few remarks. First, the structure in (6) is a general structure in the sense that predistorter (6) is derived without specific assumption on the models or characteristics of HPA, and many well known predistorters in the literature derived under specific assumptions on HPA models turn out to be some particular cases of (6). Secondly, the functions $P_{kq}$ in (6) can be computed adaptively in which, for example, the basis functions can be a mixture of polynomials, sinusoidal functions and LUTs. The adaptive selection of basis functions is being further investigated. Thirdly, when polynomials are used for each $P_{kq}$, the resulting predistorter, e.g. (10), is a reduced form of full Volterra model. This in fact can be viewed as an independent justification of the full Volterra model because (10) says that it is reasonable to use a finite Volterra series as the predistorter even if we don't make the same assumption on the HPA. Structure (10) can also be viewed as a novel and effective way of simplifying the full Volterra model. Last, but not the least, when the model and characteristics of the HPA are known, it is possible to find good predistorters by exploring the relationship between the HPA and its inverse as in [14], [17] and [18], and deriving an explicit form for the inverse. The approach taken in this paper is to design a predistorter simply relying on the observations of input and output signals of HPAs regardless of how the HPAs are made up.

## IV. SIMULATIONS AND LAB EXPERIMENTS

In this section, we present some simulation and lab test results.

### 4.1 Simulations

The simulations are set up according to Figure 1. The predistorter used in the simulations is given by (11) with $Q = 3$, and it is given by

$$\begin{aligned}z_n &= P_r(x_n, x_{n-1}, x_{n-2}) \\ &= x_n P_{11}(|x_n|) P_{12}(|x_{n-1}|) P_{13}(|x_{n-2}|) \\ &\quad + x_{n-1} P_{21}(|x_n|) P_{22}(|x_{n-1}|) P_{23}(|x_{n-2}|) \\ &\quad + x_{n-2} P_{31}(|x_n|) P_{32}(|x_{n-1}|) P_{33}(|x_{n-2}|).\end{aligned} \quad (13)$$

Each of the functions $P_{kq}$ is a polynomial of degree $M - 1 = 4$. The reason why polynomials of degree 4 are chosen is to limit the complexity while meeting the performance. The highest degree in (13) is odd when the degree of $P_{kq}$ is even. Although it is well known that only odd terms are present in polynomial or Volterra HPA models, the inverse of HPA may no longer be represented by polynomials, finite Volterra series or the exact form of (13). Therefore, the approximation of the inverse of HPA may well contain even terms. Another way to put it is that the inverse of HPA can in general be approximated better using polynomials containing both even and odd terms in (13) than using polynomials of odd degree only, because no restriction is placed on the values of the coefficients. The polynomial predistorter of [2] contains also even terms.

The polynomials are expressed in terms of an orthonormal basis $\{\psi_0(|x|), ..., \psi_{M-1}(|x|)\}$, i.e.,

$$P_{kq}(|x|) = \sum_{m=0}^{4} u_m^{kq} \psi_m(|x|), \quad (14)$$

where $\psi_m(|x|)$ is an orthogonal polynomial of degree $m$ whose weight function is the density function of the transmitted signal. The use of orthogonal polynomials for predistortion was first suggested in [15] and [16]. The orthonormal basis functions are defined as

$$\int_0^1 \rho(x)x^2\psi_i(x)\psi_j(x)\,dx = \begin{cases} 0 & i \neq j \\ 1 & i = j, \end{cases}$$

where the density function $\rho(x)$ is estimated by using the histogram method; see [12]. The orthonormal polynomials can be computed using a three term recursion.

For the purpose of plotting the predistorter functions $P_{kq}$, these functions will be scaled so that their norms are equal to 1, i.e., $\|P_{kq}\| = 1$, because without scaling, it is difficult to view the overall shape of these functions when they have different magnitudes. With the scaling, we can rewrite (13) as

$$\begin{aligned} z_n &= P_r(x_n, x_{n-1}, x_{n-2}) \\ &= a_1 x_n P_{11}(|x_n|)P_{12}(|x_{n-1}|)P_{13}(|x_{n-2}|) \\ &\quad + a_2 x_{n-1} P_{21}(|x_n|)P_{22}(|x_{n-1}|)P_{23}(|x_{n-2}|) \\ &\quad + a_3 x_{n-2} P_{31}(|x_n|)P_{32}(|x_{n-1}|)P_{33}(|x_{n-2}|), \end{aligned} \quad (15)$$

where $a_k, k = 1, 2, 3$ are scaling factors.

An OFDM signal as specified in DVB-SH with 16QAM modulation is used in the simulations. The functions $P_{kq}$ are computed in the adaptation block of Figure 1 by using the stochastic conjugate gradient (SCG) method of [12] which solves the minimization problem of (8). Instead of computing the coefficients of $u_m^{kq}$ in (14), the SCG method computes the functions directly, in the form of LUTs. In the adaptation block, samples from the input and output of the HPA are taken and they are used to estimate the predistorter functions $P_{kq}$. At each iteration, a total of 25,600 samples are taken for each of $y_n, z_n$, where $y = F(z)$.

Two HPA models will be used for simulations. In the first simulation, the HPA model is the memory polynomial model given in Example 2 of [2]. That is, the operator $F$ is defined as

$$y_n = F(z) = \sum_{\substack{k=1 \\ k\ odd}}^{5} \sum_{q=0}^{2} c_{kq} z_{n-q} |z_{n-q}|^{k-1}. \quad (16)$$

The coefficients $c_{kq}, k = 1, 3, 5, q = 0, 1, 2$ are complex numbers and their values can be found in [2, Equation (12)]. This model was chosen because it represents an actual Class AB HPA [2], is well studied, easy to implement, and the simulation results using the model have been known to correlate well with lab experiments.

After the SCG method converges, the estimate of the predistorter functions $P_{kq}$ are obtained. The results are shown in Figure 6.

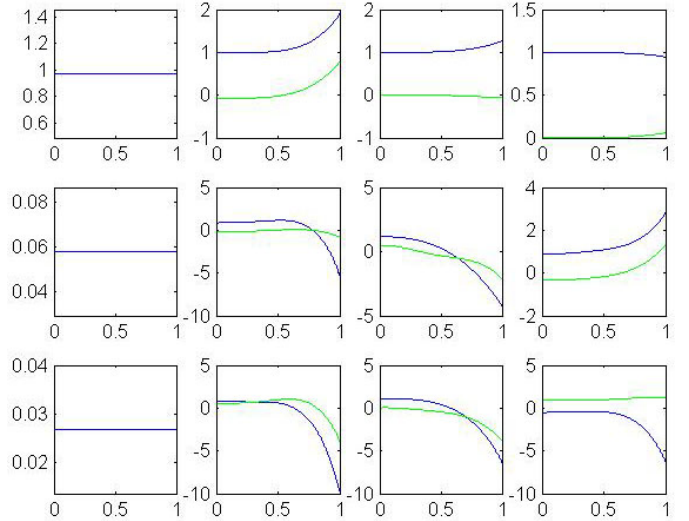

Figure 6. Computed predistorter functions for HPA model (16)

There are $3 \times 4$ plots in Figure 6, and they represent the scaling factors $a_k$ and $P_{kq}$. The first column shows the three scaling factors, and columns 2 through 4 show the scaled functions $P_{kq}$. For example, the four plots in the first row represent $a_1$, $P_{11}, P_{12}$, and $P_{13}$. Similarly, the third row represents $a_3$, $P_{31}, P_{32}$, and $P_{33}$.

After the predistorter matrix is computed, it is used in the predistorter block of Figure 1. Linearization is achieved with the cascade of predistorter and HPA. This is evident by observing the spectrum of the HPA signal when the predistorter is applied. The spectra of the signals with and without predistortion are shown in Figure 7. The red curve, which is the the top curve having the highest shoulder values, is the spectrum of the HPA output signal without using a digital predistortion. The green curve, the middle curve, is the spectrum of the HPA output signal when the predistorter is applied. The blue curve, the bottom one, is the spectrum of the original input signal.

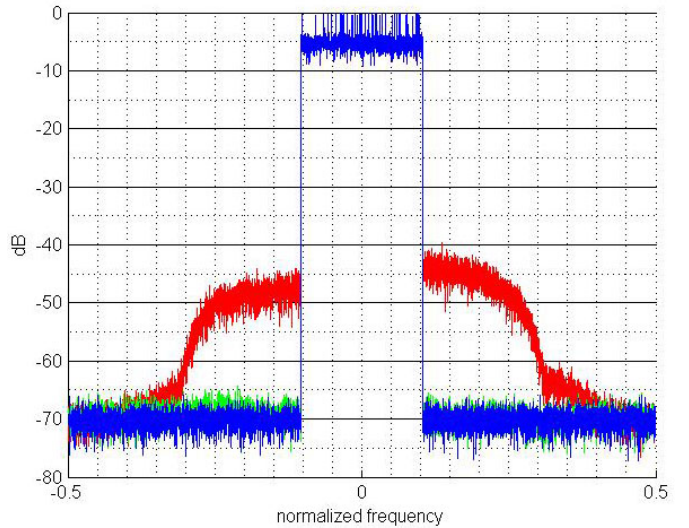

Figure 7. Spectra of signals using model (16) with predistorter (13). From top to bottom: no predistortion, with predistortion (13), original input signal

As is evident from Figure 7, the predistorter of (13), or equivalently (15), is quite effective in compensating the nonlinear effects of the HPA model of (16) because, with the predistorter, the HPA output signal is almost identical to the original input signal.

It is worthwhile to point out that plots similar to Figure 7 are obtained when the predistorters of [1] and [2] are used. In other words, although the predistorter of [2] is only a simplified case of (13), it is adequate enough to handle the HPA model of (16). Also, the predistorter of form (12) has the same performance as shown in Figure 7. We further point out that predistorter (13) has the similar performance for other HPA models given in [2].

In another simulation, we modify the HPA model of (16). A cross term is added as

$$y_n = F(z) = \sum_{\substack{k=1 \\ k\,odd}}^{5} \sum_{q=0}^{2} c_{kq} z_{n-q} |z_{n-q}|^{k-1} \quad (17)$$
$$+ 0.5 z_{n-2} |z_{n-1}| |z_n|.$$

In (17), the coefficients $c_{kq}$ have the same values as defined in (16) or [2, Equation (12)]. The HPA model of (17) is a simple extension of (16), falling into the general case of memory polynomial models, with the additional cross term to model effects of HPA with very wideband signals such as multicarrier waveforms. For this HPA model, the computed scale factors and predistorter functions, $a_k$ and $P_{kq}$, are shown in Figure 8.

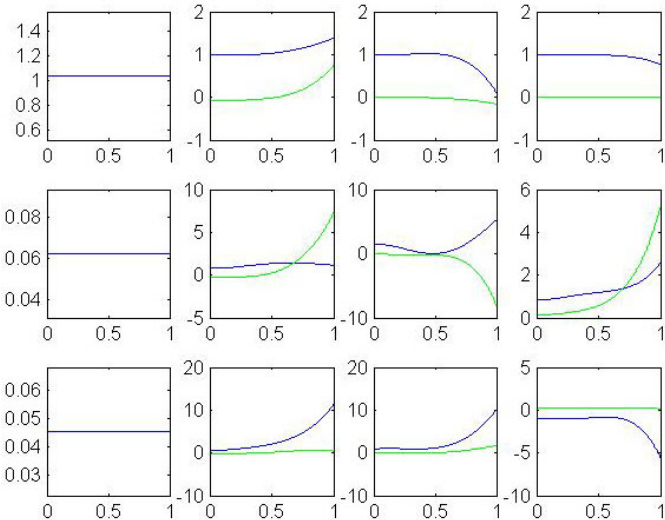

Figure 8. Computed predistorter functions for HPA model (17)

The spectra of the signals with HPA model (17) are shown in Figure 9. As is evident from Figure 9, the predistorter of (13) is quite effective in compensating the nonlinear effects of the HPA model of (17), since the spectrum of the HPA output signal with the predistorter is almost indistinguishable from that of the original input signal.

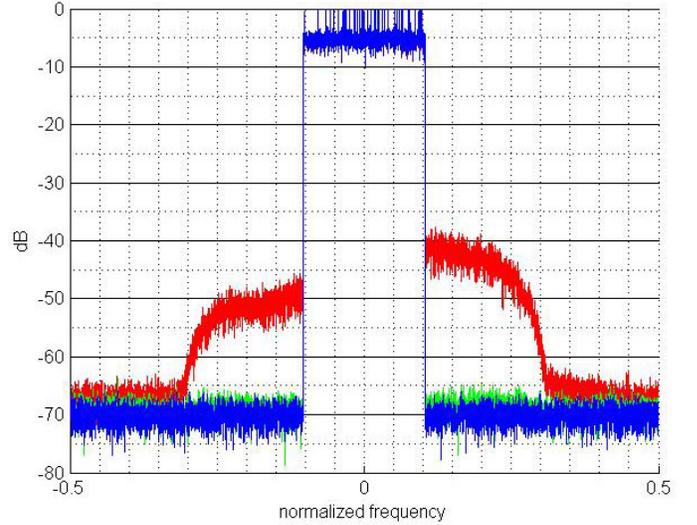

Figure 9. Spectra of signals in HPA model (17) with predistorter (13). From top to bottom: no predistorter, with separable function predistorter (13), the original input signal.

As a comparison, we also ran a simulation with the predistorter of from (12). The same number of functions as in (13) is used. The predistorter is given by

$$\begin{aligned} z_n &= P_r(x_n, x_{n-1}, x_{n-2}) \\ &= x_n [P_{11}(|x_n|) + P_{12}(|x_{n-1}|) + P_{13}(|x_{n-2}|)] \\ &+ x_{n-1} [P_{21}(|x_n|) + P_{22}(|x_{n-1}|) + P_{23}(|x_{n-2}|)] \\ &+ x_{n-2} [P_{31}(|x_n|) + P_{32}(|x_{n-1}|) + P_{33}(|x_{n-2}|)], \end{aligned} \quad (18)$$

where each $P_{kq}$ is a polynomial of degree 4. The spectra of the signals when predistorter (18) is used are shown in Figure 10.

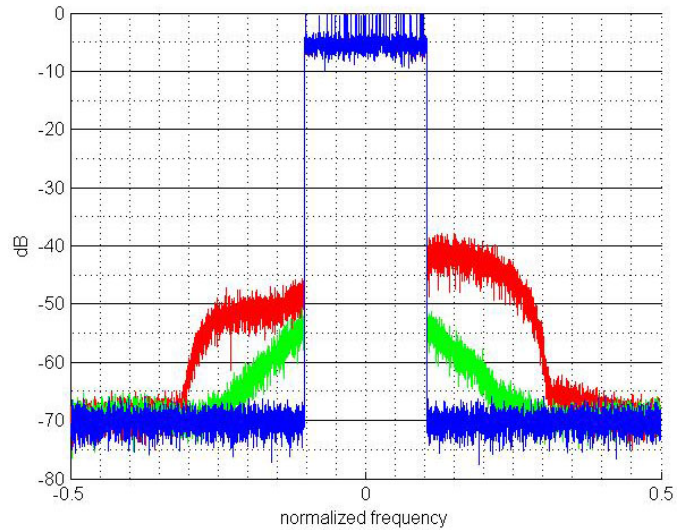

Figure 10. Spectra of signals in HPA model (17) with predistorter (18). From top to bottom: no predistorter, with predistorter (18), original input signal.

A comparison of Figure 9 and Figure 10 shows that the separable function predistorter (13) is much more effective than (18). Both (13) and (18) have a similar complexity, except for the six complex additions in (18) as opposed to the six complex multiplications in (13). However, while the linearization is achieved by predistorter (13), predistorter (18) is not able to linearize HPA of model (17).

Note that the generalized memory polynomial of [1] is a special case of (18). These results demonstrate the strength of the separable function predistorter of form (11): it is a systematic and general predistorter structure capable of compensating the nonlinearity for a variety of HPA models.

In Matlab simulations, each iteration to update the predistorter (13) takes about 5 seconds on a Gateway laptop with an Intel T5450 1.66GHz processor. It takes about 20 iterations for convergence.

### 4.2 Lab Experiments

Preliminary lab test was performed on a Class B power amplifier with average output power of 56W using the UMTS waveform. The predistorter (15) was implemented in a high speed Field Programmable Gate Array (FPGA) board. Each function $P_{kq}$ is implemented as an LUT in the FPGA. The result of the preliminary test is shown in Figure 11, in which spectrum analyzer traces of spectra of HPA output for four carrier UMTS signal are shown. In Figure 11 the curve in black color, which is the curve with higher spectrum regrowth, is the spectrum of the HPA output without predistortion, and the blue curve, the one with lower regrowth, is the spectrum of the HPA output with the predistortion. As shown, the spectrum regrowth is suppressed by about 20dB after predistortion, and the adjacent channel power (ACP) after predistortion is about 55dB below the carrier power, which demonstrates that the predistortion method of this paper is very effective.

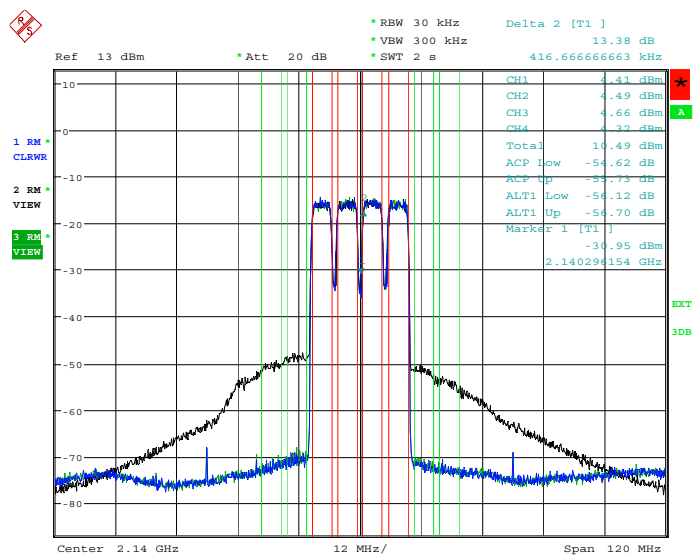

Figure 11. Spectra of HPA output for four carrier UMTS signal with and without predistortion

The spurs in the signal with predistortion at +/-30MHz from the center frequency were caused by interference on the test board, and they are not related to the predistortion.

The computed scale factors and predistorter functions, $a_k$ and $P_{kq}$, after 40 iterations are shown in Figure 12.

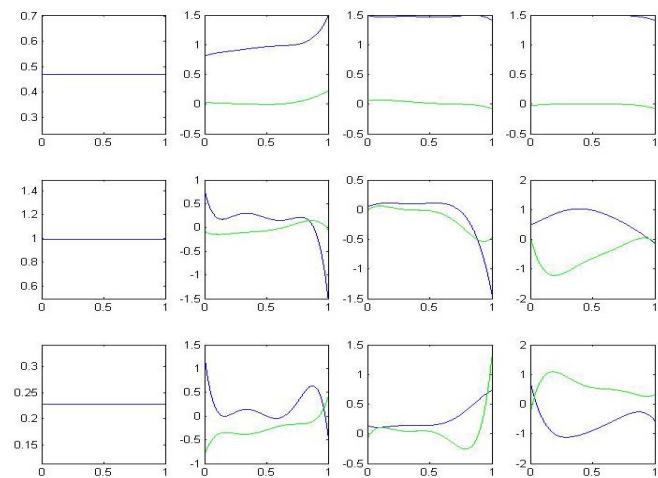

Figure 12. Computed predistorter functions for Class B power amplifier with four carrier UMTS waveform

In the preliminary lab test, although the LUTs $P_{kq}$ themselves are implemented in FPGA, the computation of the LUTs is carried out "offline" where a set of input/output samples is captured from HPA, and the samples are transferred to a PC where entries of LUTs are computed, and transferred into FPGA. In this setting, each iteration, consisting of capturing samples, transferring the data, computing the entries, updating the FPGA, takes about 2 to 3 minutes, with the majority of time being spent on capturing and transferring the input/output samples. While this speed of computation is adequate for broadcast systems such as ATSC, DVB-T/H/SH due to their constant transmission power, improvement is needed for wireless communication systems in which transmission power may vary in time. When an iterative method such as the stochastic conjugate gradient (SCG) method [12] is used, the computation of $P_{kq}$ can be easily implemented in FPGA to meet the requirement of wireless communication systems.

More experiments are under way for other types of HPAs, with waveforms of different bandwidths, using predistorters with non-polynomial basis functions for $P_{kq}$, and predistorter of structure in (6) and Figure 3. The results will be presented in another paper.

### V. CONCLUSIONS

We have shown that under certain conditions, the predistorter and the postdistorter are equivalent. Subsequently, by using approximation by separable functions, we have established a systematic and general structure for a predistorter that is capable of compensating nonlinearity for a large variety of HPA models. This systematic structure avoids the selection of predistorters according the assumption of HPA models traditionally done in the literature. Simulations have been performed to verify this work.

## ACKNOWLEDGEMENT

The authors would like to thank Dennis Morgan and two anonymous reviewers for their careful reading of this paper and their valuable comments leading to improvement of the paper. The authors are also grateful to Xin Yu of Alcatel-Lucent who performed the preliminary lab tests.